\documentclass[preprint,preprintnumbers,amsmath,amssymb]{revtex4}
\usepackage{amssymb}
\usepackage{latexsym}
\usepackage{epsfig}
\usepackage{color}
\allowdisplaybreaks
\begin{document}
\title{ Thermodynamics in Rastall Gravity with Entropy Corrections}
\author{Kazuharu Bamba$^1$\footnote{bamba@sss.fukushima-u.ac.jp},
Abdul Jawad$^2$\footnote{jawadab181@yahoo.com;
abduljawad@cuilahore.edu.pk}, Salman
Rafique$^2$\footnote{salmanmath004@gmail.com}, Hooman
Moradpour$^3$\footnote{h.moradpour@riaam.ac.ir}}
\address{$^1$ Division of Human Support System, Faculty of Symbiotic Systems Science, Fukushima University, Fukushima 960-1296, Japan\\
$^2$ Department of Mathematics, COMSATS University Islamabad, Lahore Campus, Lahore-54000, Pakistan\\
$^3$ Research Institute for Astronomy and Astrophysics of Maragha
(RIAAM), Maragha 55134-441, Iran}
\date{}

\begin{abstract}
We explore the thermodynamic analysis at the apparent horizon in the
framework of Rastall theory of gravity. We take different entropies
such as the Bakenstein, logarithmic corrected, power law corrected,
and the Renyi entropies. We investigate the first law and
generalized second law of thermodynamics analytically for these
entropies which hold under certain conditions. Furthermore, the
behavior of the total entropy in each case is analyzed. As a result,
it is implied that the generalized second law of thermodynamics is
satisfied. We also check whether the thermodynamic equilibrium
condition for these entropies is met at the present horizon.

\noindent\textbf{Keywords:Thermodynamics, Rastall gravity, Entropy.}\\
\textbf{PACS:} 95.36.+d; 98.80.-k \\
{\bf Report number:} FU-PCG-24
\end{abstract}
\maketitle

\section{Introduction}

The universality of the conservation law of energy and momentum,
$T^{\mu\nu};_\mu=0$, where $T^{\mu\nu}$ is the energy-momentum
tensor, in both flat and the curved spacetimes is one of the
Einstein's basic assumptions to get general relativity \cite{s1,
s2}. With the help of this generalization to formulate the Mach
principle, Einstein has obtained his famous tensor and then related
field equations leading to the second order equation of motion
\cite{s1, s2} which have too many applications in astrophysics and
cosmology \cite{s2, s3}. In 1972, by relating $T^{\mu\nu};_\mu$ to
the derivative of the Ricci scalar, Rastall proposed a new
formulation for gravity which converges to the Einstein formulation
in the flat background (empty universe) \cite{1}. Indeed, he argued
that the $T^{\mu\nu};_\mu=0$ assumption made by Einstein to obtain
his field equations, is questionable in the curved spacetimes
\cite{1}. 
In fact, for $T^{\mu\nu};_\mu\neq0$, the gravitationally induced
particle creation in cosmology is phenomenologically confirmed
\cite{s3b}--\cite{s3d}. Moreover, in a gravitational system, quantum
effects lead to the violation of the condition $T^{\mu\nu};_\mu=0$
\cite{s4d}. Hence, $T^{\mu\nu};_\mu$ is directly related with the
Ricci scalar, and therefore the Rastall theory may be considered as
a classical formulation for the particle creation in cosmology
\cite{s5d}. In order to explain the issues regarding late-time
cosmic acceleration, different dark energy models and modified
theories of gravity has been presented, see, for instance,
\cite{R-DE-MG}-\cite{Nojiri:2017ncd}.

After numerous years in the time of Einstein, Jacobson \cite{s4}
demonstrated that one would be able to acquire the Einstein
equations with the help of the Clausius relation on the local
Rindler causal horizon. Actually, the purpose of the Jacobson's work
is for spacetimes with a causal horizon that the Einstein equations
would be considered as a thermodynamical equation of state on the
horizon, if one generalizes the four law of black holes to the
causal horizon. Furthermore, Eling et al. \cite{s5} demonstrated
that terms other than the Einstein-Hilbert, one can produce entropy
due to non-equilibrium thermodynamic aspects to generalized $f(R)$
theory by the jacobson's idea, which yields the modification of the
event horizon entropy \cite{s5,s6}. In fact, applying the
thermodynamics laws to the horizon, and using the field equations,
one can find the horizon entropy in various cosmological and
gravitational setups
\cite{s6,s7,s8,s9,s10,s12,s13,j6,j7,j13,msgj,plb,ms,plb1c}.

The generalized second law of thermodynamics (GSLT) has also been
studied extensively in the behavior of expanding universe. According
to GSLT, ``\textit{the entropy of matter inside the horizon plus
entropy of the horizon remains positive and increases with the
passage of time}'' \cite{s14}. It is assumed that the horizon
entropy is given by the quarter of its area \cite{s12} or power law
correction \cite{s16}-\cite{s162} or logarithmic entropy \cite{s17}
and the Reyni entropy to analyze the validity of GSLT.
Thermodynamics of a Schwarzschild black hole in phantom cosmology
with entropy corrections has also been examined \cite{Bamba:2012mj}.
Most of the researchers have discussed the validity of GSLT of
different system including the interaction of two fluid components,
dark energy (DE) and dark matter \cite{s18}-\cite{s183}, and that of
three fluid components (DE, dark matter and radiation)
\cite{s19}-\cite{s192} in the FRW universe. Cosmological
investigations of thermodynamics in modified gravity theories have
been executed in Refs.~\cite{CT}-\cite{CT6} (for a recent review on
thermodynamic properties of modified gravity theories, see,
e.g.,~\cite{Bamba:2016aoo}).

Recently, applying the thermodynamics laws to the spacetime horizon
and using the Rastall field equations, the horizon entropy has been
obtained in both the static and dynamic setup \cite{plb,ms,plb1c}.
These results show that the horizon entropy in the Rastall theory
differs from that of the Einstein theory, a signal addressing us
that their Lagrangian are also different
\cite{lag1,lag2,lag3,lag4,epjcc0,epjcc}. In addition, it has also
been shown that the R\'{e}nyi entropy content of horizon can help us
in providing a proper description for the current accelerated
universe in both the Einstein and Rastall theory \cite{non20}, an
analysis which also reveals some differences between the
cosmological features of the Rastall theory and those of the
Einstein theory. It is also useful to mention here that the Rastall
theory provides a proper platform for generalizing the unimodular
gravity which leads to the interesting cosmological consequences
\cite{fabun}. Some authors have also given their analysis on Rastall
theory \cite{Visser:2017gpz,Moradpour:2017tbp}.

In this paper, our aim is to discuss the validity of first law of
thermodynamics, GSLT and thermodynamical equilibrium of the FRW
universe in the Rastall theory of gravity in the presence of the
equation of state (EoS) $p=\rho(\gamma-1)$ (where $p$ is the
pressure, $\rho$ is the energy density and $\gamma$ is a EoS
parameter). By applying the Clausius relation on the apparent
horizon of the FRW universe, we get the validity of first law of
thermodynamics in different entropy corrections. We also analyze the
validity of GSLT and thermodynamical equilibrium on apparent horizon
by assuming the different entropies such as Bekenstein entropy,
logarithmic corrected entropy, power law corrected entropy and the
Renyi entropy in Rastall theory of gravity.

The scheme of this paper is organized as follows. In section 2, we
present the basic equations, Rastall theory and cosmological
parameters. In Section 3, we discuss thermodynamics on the apparent
horizon using Bekenstein entropy. We investigate logarithmic
corrected entropy, power law corrected entropy and the Renyi entropy
in sections 4, 5 and 6 respectively. Finally, conclusions are given
in Section 7.

\section{Basic Equations}

On the basis of Rastall theory of gravity, the ordinary
energy-momentum conservation law is not always available in the
curved spacetime and therefore we should have
\begin{eqnarray}\label{E1}
T^{\mu\nu};_\mu=\lambda R^{,~\nu},
\end{eqnarray}
where $R$ and $\lambda$ are the Ricci scalar of the spacetime and
the Rastall constant parameter respectively which should be
determined from observations and other parts of physics \cite{1}.
With the help of above relation, a generalization of the
gravitational field can be found as
\begin{eqnarray}\label{E2}
G_{\mu\nu}+k\lambda g_{\mu\nu}R=kT_{\mu\nu},
\end{eqnarray}
here $G_{\mu\nu},~T_{\mu\nu}$ and $k$ are Einstein tensor,
energy-momentum tensor and coupling constant respectively. Moreover
for $\lambda=0$, the Einstein field equations can be re-covered
\cite{1}. The line element of FRW universe can be written as
\begin{eqnarray}\label{E3}
ds^2=-dt^2+a^2(t)\bigg(\frac{dr^2}{1-\kappa
r^2}+r^2\big(d\theta^2+\sin \theta^2d\phi^2\big)\bigg).
\end{eqnarray}
In this equation $a(t)$ and $\kappa$ are scale factor and curvature
parameter respectively, while $\kappa=-1, 0, 1$ denotes the open,
flat and closed universe respectively \cite{3}. We consider the
$\kappa=0$ for flat universe for which Freidmann equations in
Rastall thoery can be be obtained by using Eqs.(\ref{E2}) and
(\ref{E3}) as
\begin{eqnarray}\label{E4}
(12k\lambda-3)H^2+6k\lambda\dot{H}&=&-k\rho \\\label{E5}
(12k\lambda-3)H^2+(6k\lambda-2)\dot{H}&=&-kp,
\end{eqnarray}
where $\rho$ is energy density and $p$ is pressure of
energy-momentum source.

The Bianchi identity implies $G^{; \mu}_{\mu\nu}=0$ which leads to
the equation of continuity \cite{4} as follows
\begin{eqnarray}\label{E6}
\frac{3k\lambda-1}{4k\lambda-1}\dot{\rho}+\frac{3k\lambda}{4k\lambda-1}\dot{p}
+3H\big(\rho+p\big)=0.
\end{eqnarray}
From above equation, one can rediscover the Friedmann equations and
equation of continuity by taking $\lambda=0$ and $k=8\pi$. Further,
combining Eqs.(\ref{E4}) and (\ref{E5}) and applying EoS parameter
$p=(\gamma-1)\rho$ where $\frac{2}{3}\leq\gamma\leq 2$, we get
\begin{eqnarray}\label{E7}
\dot{H}=-\frac{k}{2}\big(\gamma\rho\big),
\end{eqnarray}
which is independent of $\lambda$. It is same as that of the
standard cosmology, which depends on the Einstein theory and the FRW
metric. Inserting the value of $\dot{H}$ in Eq.(\ref{E4}), it yields
\begin{eqnarray}\label{E8}
H^2=\frac{k\rho\big(3k\lambda
\gamma-1\big)}{3\big(4k\lambda-1\big)}.
\end{eqnarray}
Integration of Eq.(\ref{E6}) leads to the solution
$\rho=ba^{-\frac{3\gamma(4k\lambda-1)}{(3k\lambda \gamma-1)}}$. By
putting this value in Eq.(\ref{E8}), we obtain
\begin{eqnarray}\label{E10}
H=\sqrt{\frac{kb\big(3k\lambda
\gamma-1\big)a^{-\frac{3\gamma(4k\lambda-1)}{(3k\lambda
\gamma-1)}}}{3\big(4k\lambda-1\big)}}.
\end{eqnarray}
It can be observed from this equation that the Hubble parameter
becomes positive for $k>0,~b>0$ and $\lambda<\frac{1}{3\gamma k}$
(or $\gamma<\frac{1}{3\lambda k}$ which leads to the constraint
$\frac{2}{3}\leq\frac{1}{3\lambda k}\leq 2$).

In the following, we analyze the validity of first law of
thermodynamics, GSLT and thermodynamical equilibrium in the presence
of different entropies such as Bekenstein entropy, logarithmic
corrected entropy, power law corrected entropy and the Renyi
entropy.

\section{Thermodynamical Analysis for the modified Bekenstein Entropy}

Rastall gravitational field equations and Rastall Lagrangian are
different from Einstein theory \cite{5}. Therefore one can expect
that the horizon entropy is in Rastall theory differs from
Bekenstein entropy. In the flat FRW universe, apparent horizon
relates with Hubble parameter as $R_A=\frac{1}{H}$. Taking first
derivative with respect to time, we get
\begin{eqnarray}\label{E11}
 \dot{R_A}=-\frac{\dot{H}}{H^2}=\frac{k\gamma ba^{-\frac{3\gamma(4k\lambda-1)}{(3k\lambda
\gamma-1)}}}{2H^2}.
 \end{eqnarray}
However, the modified Bekenstein entropy in Rastall theory on the
apparent horizon takes the following form on \cite{hooman}
 \begin{eqnarray}\label{E120}
S_{\textmd{A}}=\frac{\tilde{A}}{4}~~\text{and}
~~\text{where}~~\tilde{A}=(1+\frac{2\gamma}{1+4\gamma})A~~with~~A=4\pi
R^2_A,
\end{eqnarray}
and the units of $c=\hbar=G=1$ has been considered. Recently, it has
been proposed that the horizon entropy in the Rastall theory is the
same as that of the Einstein theory \cite{epjcn}, a result in
contrast with the above equation. In Ref. \cite{epjcn}, authors used
the Misner-Sharp mass of the Einstein theory, but in Ref.
\cite{hooman}, the Misner-Sharp mass of the Rastall theory is used
to obtain the horizon entropy. Since the Misner-Sharp mass depends
on the gravitational theory under investigation \cite{mis}, we take
into account Eq.~(\ref{E120}) as the horizon entropy in agreement
with others attempts \cite{clas}. Also, the Hawking temperature at
apparent can be defined as \cite{CaiKimt}
\begin{eqnarray}\label{E12}
T_{\textmd{A}}=\frac{1}{2\pi R_A}
\end{eqnarray}

The differential $dE_{\textmd{A}}$ is the amount of energy
crossing the apparent horizon can be evaluated as \cite{6}
\begin{eqnarray}\label{E14}
-dE_{\textmd{A}}=\frac{1}{2}R^3(\rho+p)Hdt=\frac{\gamma
ba^{-\frac{3\gamma(4k\lambda-1)}{(3k\lambda \gamma-1)}}}{2H^2}dt.
\end{eqnarray}
From Eq.(\ref{E12}) we can get the differential of surface entropy
which leads to
\begin{eqnarray}\label{E16}
T_AdS_A=\frac{(1+\frac{2\gamma}{1+4\gamma})k\gamma
ba^{-\frac{3\gamma(4k\lambda-1)}{(3k\lambda \gamma-1)}}}{2H^2}dt.
\end{eqnarray}
The first law of thermodynamics is given with the help of the
Clausius relation $-dE_{A}=T_{A}dS_{A}$ written as
\begin{eqnarray}\label{E13a}
\Omega\ dt=T_{A}dS_{A}+dE_{A},
\end{eqnarray}
for the sake convenience, which leads to
\begin{eqnarray}\label{E16a}
\Omega=\frac{(1+\frac{2\gamma}{1+4\gamma})k\gamma
ba^{-\frac{3\gamma(4k\lambda-1)}{(3k\lambda
\gamma-1)}}}{2H^2}-\frac{\gamma
ba^{-\frac{3\gamma(4k\lambda-1)}{(3k\lambda \gamma-1)}}}{2H^2}.
\end{eqnarray}
Therefore, the first law of thermodynamics holds when
$\Omega\rightarrow 0$ which leads to a constraint $\gamma=
\frac{1-k}{2(2k-1)}$.

Now we check the validity of GSLT and thermodynamical equilibrium
for an isolated macroscopic physical system having maximum entropy
state. Second law of thermodynamics has been generalized towards the
cosmological system where it can be defined as the sum of all
entropies of the constituents (mainly dark matter and DE) and
entropy of boundary (either it is Hubble or apparent or event
horizons) of the universe can never decrease, i.e.,
$d(S_{\textmd{A}}+S_{\textmd{f}})\geq0$. The Gibbs equation is of
the form
\begin{eqnarray}\label{E17}
T_fdS_f=dE_f+p dV,
\end{eqnarray}
where $ T_{\textmd{f}} $ is the temperature of the cosmic fluid and
$ E_{\textmd{f}}$ is the energy of the fluid ($E_{\textmd{f}}=\rho
V) $. From Eq.(\ref{E17}) we can find the differential of fluid
entropy as
\begin{eqnarray}\label{E18}
dS_f=\frac{\gamma b}{8H^3}a^{-\frac{3\gamma(4k\lambda-1)}{(3k\lambda
\gamma-1)}}\bigg(\frac{(1-4k\lambda)}{(3k\lambda\gamma-1)}+\frac{k\gamma
b}{2H^2}a^{-\frac{3\gamma(4k\lambda-1)}{(3k\lambda
\gamma-1)}}\bigg)dt.
\end{eqnarray}
The total rate of change of entropy is given by
\begin{eqnarray}\label{E19}
\dot{S_T}=\frac{\gamma ba^{-\frac{3\gamma(4k\lambda-1)}{(3k\lambda
\gamma-1)}}}{8H^3}\bigg(\frac{(1-4k\lambda)}{(3k\lambda\gamma-1)}+\frac{k\gamma
b}{2H^2}a^{-\frac{3\gamma(4k\lambda-1)}{(3k\lambda
\gamma-1)}}\bigg)+\frac{(1+\frac{2\gamma}{1+4\gamma})k\gamma
ba^{-\frac{3\gamma(4k\lambda-1)}{(3k\lambda \gamma-1)}}}{8H^3}.
\end{eqnarray}
For the validity of GSLT, $\dot{S_T}\geq0$ which gives us the
following relation of scale factor
\begin{eqnarray}\nonumber
a\geq\bigg(\frac{t}{\sqrt{Q}}+D\bigg),
\end{eqnarray}
here $D$ is the integration constant,
$Q=\frac{4k\lambda-1}{3k\lambda\gamma-1}-(1+\frac{2\gamma}{4\gamma+1})$
and $P=\frac{-3\gamma(4k\lambda-1)}{3k\lambda\gamma-1}$. Taking the
expression $\acute{S}_T=\frac{dS_T}{da}=\frac{\dot{S_T}}{a H}$ and
replacing the value of $\dot{S_T}$, this equation becomes
\begin{eqnarray}\label{E19b}
{S'}_T&=&\bigg(\frac{\gamma
ba^{-\frac{3\gamma(4k\lambda-1)}{(3k\lambda
\gamma-1)}}}{8H^3}\bigg(\frac{(1-4k\lambda)}{(3k\lambda\gamma-1)}+\frac{k\gamma
ba^{-\frac{3\gamma(4k\lambda-1)}{(3k\lambda
\gamma-1)}}}{2H^2}\bigg)\nonumber\\
&+&\frac{(1+\frac{2\gamma}{1+4\gamma})k\gamma
b}{8H^3}a^{-\frac{3\gamma(4k\lambda-1)}{(3k\lambda
\gamma-1)}}\bigg)\frac{1}{a H}.
\end{eqnarray}
\begin{figure} \centering
\epsfig{file=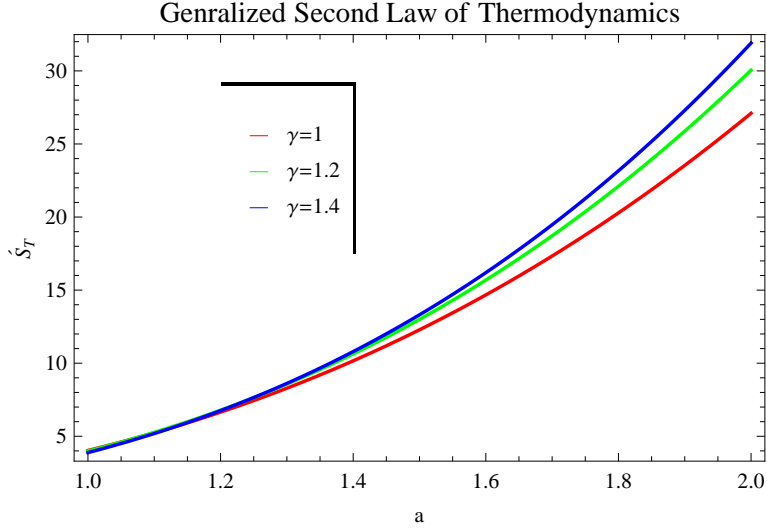,width=.70\linewidth}\caption{Plot of
${S'}_T$ versus $a$ for Bekenstein entropy using $k=1,~\lambda=-1$
and $b=1$.}
\end{figure}
The graphical behavior of ${S'}_T$ versus scale factor $a$ is shown
in Figure \textbf{1}. It can be observed that GSLT satisfy the
condition ${S'}_T\geq0$ for chosen three values of $\gamma$ which
leads to the validity of GSLT.

In order to discuss the thermodynamical equilibrium, we obtain the
second order differential equation by using Eq.(\ref{E19b}), as
follows
\begin{eqnarray}\label{E19d}
S^{''}_T&=&\frac{3bk\gamma\sqrt{3}a^{-2-\frac{3\gamma(4k\lambda-1)}{(3k\lambda
\gamma-1)}}}{2\big(\frac{bk(3k\lambda\gamma-1)a^{-\frac{3\gamma(4k\lambda-1)}{(3k\lambda
\gamma-1)}}}{(4k\lambda-1)}\big)^\frac{3}{2}}\bigg(-\frac{9\sqrt{3}b\gamma^2(4k\lambda-1)^2
a^{-\frac{3\gamma(4k\lambda-1)}{(3k\lambda
\gamma-1)}}}{16(3k\lambda\gamma-1)^2\big(\frac{bk(3k\lambda\gamma-1)a^{-\frac{3\gamma(4k\lambda-1)}{(3k\lambda
\gamma-1)}}}{(4k\lambda-1)}\big)^\frac{3}{2}}+3bk\gamma\nonumber\\
&\times&\frac{\sqrt{3}a^{-\frac{3\gamma(4k\lambda-1)}{(3k\lambda
\gamma-1)}}\big(1+\frac{2\gamma}{4\gamma-1}\big)}{8\big(\frac{bk(3k\lambda\gamma-1)a^{-\frac{3\gamma(4k\lambda-1)}{(3k\lambda
\gamma-1)}}}{(4k\lambda-1)}\big)^\frac{3}{2}}\bigg)-\frac{\sqrt{3}}
{a^2\sqrt{\frac{bk(3k\lambda\gamma-1)a^{-\frac{3\gamma(4k\lambda-1)}{(3k\lambda
\gamma-1)}}}{(4k\lambda-1)}}}\bigg(-\frac{(4k\lambda-1)^2}{(3k\lambda\gamma-1)^2}\nonumber\\
&\times&\bigg(\frac{9\sqrt{3}b\gamma^2a^{-\frac{3\gamma(4k\lambda-1)}{(3k\lambda
\gamma-1)}}}{16\big(\frac{bk(3k\lambda\gamma-1)a^{-\frac{3\gamma(4k\lambda-1)}{(3k\lambda
\gamma-1)}}}{(4k\lambda-1)}\big)^\frac{3}{2}}\bigg)+\frac{3bk\gamma\sqrt{3}a^{-\frac{3\gamma(4k\lambda-1)}{(3k\lambda
\gamma-1)}}\big(1+\frac{2\gamma}{4\gamma-1}\big)}{8\big(\frac{bk(3k\lambda\gamma-1)a^{-\frac{3\gamma(4k\lambda-1)}{(3k\lambda
\gamma-1)}}}{(4k\lambda-1)}\big)^\frac{3}{2}}\bigg)+\frac{\sqrt{3}}{a}\nonumber\\
&\times&\frac{1}{\sqrt{\frac{bk(3k\lambda\gamma-1)a^{-\frac{3\gamma(4k\lambda-1)}{(3k\lambda
\gamma-1)}}}{(4k\lambda-1)}}}\bigg(\frac{27b^2k^2\gamma^2a^{-1-\frac{6\gamma(4k\lambda-1)}{(3k\lambda
\gamma-1)}}\big(1+\frac{2\gamma}{4\gamma-1}\big)}{16\big(\frac{bk(3k\lambda\gamma-1)a^{-\frac{3\gamma(4k\lambda-1)}{(3k\lambda
\gamma-1)}}}{(4k\lambda-1)}\big)^\frac{5}{2}}-\frac{81b^2k\gamma^3\sqrt{3}}{32(3k\lambda\gamma-1)^2}\nonumber\\
&\times&\frac{a^{-1-\frac{6\gamma(4k\lambda-1)}{(3k\lambda
\gamma-1)}}(4k\lambda-1)^2}{\big(\frac{bk(3k\lambda\gamma-1)a^{-\frac{3\gamma(4k\lambda-1)}{(3k\lambda
\gamma-1)}}}{(4k\lambda-1)}\big)^\frac{5}{2}}+\frac{27b\gamma^3(4k\lambda-1)^3\sqrt{3}a^{-1-\frac{3\gamma(4k\lambda-1)}{(3k\lambda
\gamma-1)}}}{16(3k\lambda\gamma-1)^3\big(\frac{bk(3k\lambda\gamma-1)a^{-\frac{3\gamma(4k\lambda-1)}{(3k\lambda
\gamma-1)}}}{(4k\lambda-1)}\big)^\frac{3}{2}}-9bk\gamma^2\nonumber\\
&\times&\frac{\sqrt{3}(4k\lambda-1)a^{-1-\frac{3\gamma(4k\lambda-1)}{(3k\lambda
\gamma-1)}}\big(1+\frac{2\gamma}{4\gamma-1}\big)
}{8(3k\lambda\gamma-1)\big(\frac{bk(3k\lambda\gamma-1)
a^{-\frac{3\gamma(4k\lambda-1)}{(3k\lambda
\gamma-1)}}}{(4k\lambda-1)}\big)^\frac{3}{2}}\bigg).
\end{eqnarray}
\begin{figure} \centering
\epsfig{file=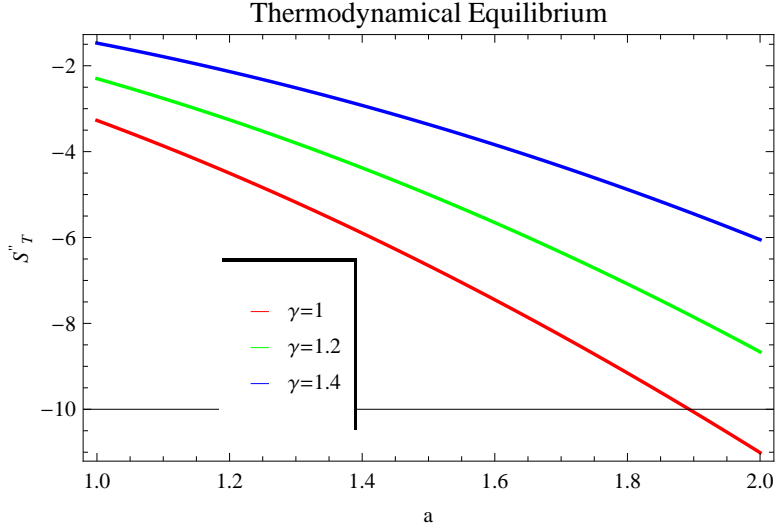,width=.70\linewidth}\caption{Plot of
$S^{''}_T$ versus $a$ for Bekenstein entropy using
$k=1,~\lambda=-1$, and $b=1$.}
\end{figure}
Figure \textbf{2} represents its plot against $a$. The trajectories
of $S^{''}_T$ indicate the positive behavior for three values of
$\gamma$. This leads to the validity of thermodynamical equilibrium
for all values of $\gamma$.

\section{Thermodynamical Analysis for Logarithmic Corrected Entropy}

To study the expansion of entropy of the universe, we discuss the
addition of entropy related to the horizon. Quantum gravity allows
the logarithmic corrections in the presence of thermal equilibrium
fluctuations and quantum fluctuations \cite{7}-\cite{epjcn1}. Using
the quantum gravity, one can get the corrected Wald entropy of
horizons as \cite{epjcn1}
\begin{eqnarray}
S=S_{W}+\alpha\ln S_{W}+....,
\end{eqnarray}
where $\alpha$ is an unknown coefficient. The attempts for the
Bekenstein-Hawking entropy ($S_{BH}$), as the Wald entropy in the
Einstein theory \cite{wald}, lead to \cite{090,0901,0902}
\begin{eqnarray}\label{ce}
S=S_{BH}+\alpha\ln S_{BH}+\frac{\beta}{S_{BH}}+...
\end{eqnarray}
where $\beta$ is constant whose value is still under consideration
(the same as $\alpha$). On one hand, Eq.~(\ref{E120}) indicates that
the difference between $S_A$, which is a proper candidate for the
Wald entropy in the Rastall theory, and $S_{BH}$ is a constant
coefficient ($1+\frac{2\gamma}{1+4\gamma}$). On the other hand, the
same result as Eq.~(\ref{ce}) is also obtainable by studying the
effects of the thermal fluctuations on the horizon entropy
\cite{90,9}, and indeed, these thermal-based approaches are not
restricted to $S_{BH}$ \cite{90,9}. Therefore, we assume
Eq.~(\ref{ce}) is also valid for the Rastall theory, and write the
logarithmic entropy corrected as
\begin{eqnarray}\label{E20}
S_A=\frac{\tilde{A}}{4L^2_p}+\alpha\ln\frac{\tilde{A}}{4L^2_p}+\beta\frac{4L^2_p}{\tilde{A}},
\end{eqnarray}
where $L_P$ is the Planck's length. The differential form of above
equation is given by
\begin{eqnarray}\label{E21}
dS_A=\frac{k\gamma ba^{-\frac{3\gamma(4k\lambda-1)}{(3k\lambda
\gamma-1)}}}{2H^2}\bigg(\frac{(1+\frac{2\gamma}{1+4\gamma})}{4HL^2_p}+2\alpha
H-\frac{16\beta H^3L^2_p}{(1+\frac{2\gamma}{1+4\gamma})}\bigg)dt,
\end{eqnarray}
which yields
\begin{eqnarray}\label{E211}
T_AdS_A=\frac{2k\gamma ba^{-\frac{3\gamma(4k\lambda-1)}{(3k\lambda
\gamma-1)}}}{H}\bigg(\frac{(1+\frac{2\gamma}{1+4\gamma})}{4HL^2_p}+2\alpha
H-\frac{16\beta H^3L^2_p}{(1+\frac{2\gamma}{1+4\gamma})}\bigg)dt.
\end{eqnarray}
Using Eq.(\ref{E13a}), we get
\begin{eqnarray}\label{E21a}
\Omega=\frac{2k\gamma ba^{-\frac{3\gamma(4k\lambda-1)}{(3k\lambda
\gamma-1)}}}{H}\bigg(\frac{(1+\frac{2\gamma}{1+4\gamma})}{4HL^2_p}+2\alpha
H-\frac{16\beta
H^3L^2_p}{(1+\frac{2\gamma}{1+4\gamma})}\bigg)-\frac{\gamma
ba^{-\frac{3\gamma(4k\lambda-1)}{(3k\lambda \gamma-1)}}}{2H^2}.
\end{eqnarray}
\begin{figure} \centering
\epsfig{file=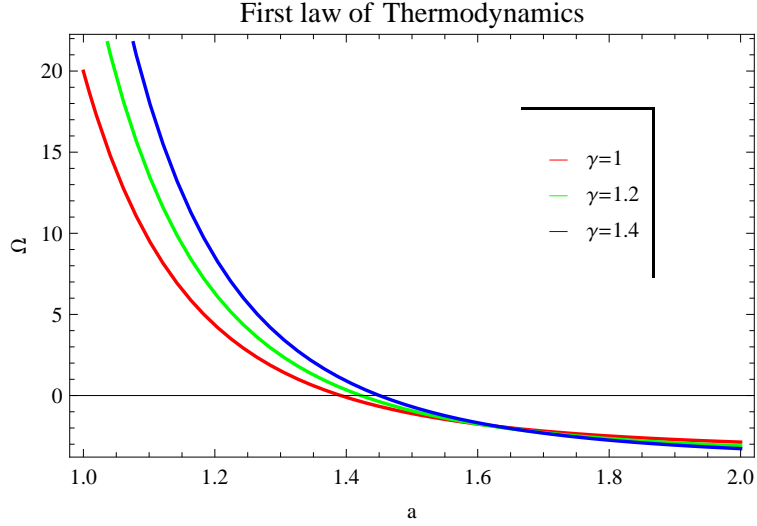,width=.70\linewidth}\caption{Plot of
$\Omega$ versus $a$ for logarithmic corrected entropy using
$k=1,~\lambda=-1,~\alpha=-1,~L_p=1,~\beta=1$ and $b=1$.}
\end{figure}
The plot of $\Omega$ versus $a$ for three values of $\gamma$ taking
same values of constants as previous case is shown in Figure
\textbf{3}. It can be observed that first law of thermodynamics
holds for some specific values of $a$ , i.e., for $\gamma=1$ at
$a=1.39$, at $a=1.43$ for $\gamma=1.2$ and for $\gamma=1.4$ at
$a=1.45$ represent the validity of first law of thermodynamics.

Moreover, we analyze the validity of GSLT and thermodynamical
equilibrium which hold if $dS_T\geq0 $ and $d^2S_T<0$ satisfy
respectively. From Eqs.(\ref{E18}) and (\ref{E21}), we get
\begin{eqnarray}\label{E22}
\dot{S_T}&=&\frac{k\gamma
ba^{-\frac{3\gamma(4k\lambda-1)}{(3k\lambda
\gamma-1)}}}{2H^2}\bigg(\frac{(1+\frac{2\gamma}{1+4\gamma})}{4HL^2_p}+2\alpha
H-\frac{16\beta
H^3L^2_p}{(1+\frac{2\gamma}{1+4\gamma})}\bigg)+\frac{\gamma
b}{8H^3}\nonumber\\
&\times&a^{-\frac{3\gamma(4k\lambda-1)}{(3k\lambda
\gamma-1)}}\bigg(\frac{(1-4k\lambda)}{(3k\lambda\gamma-1)}+\frac{k\gamma
ba^{-\frac{3\gamma(4k\lambda-1)}{(3k\lambda
\gamma-1)}}}{2H^2}\bigg).
\end{eqnarray}
This equation leads to the following equation
\begin{eqnarray}\label{E22a}
\acute{S}_T&=&\frac{1}{a H}\bigg(\frac{k\gamma
ba^{-\frac{3\gamma(4k\lambda-1)}{(3k\lambda
\gamma-1)}}}{2H^2}\bigg(\frac{(1+\frac{2\gamma}{1+4\gamma})}{4HL^2_p}+2\alpha
H-\frac{16\beta
H^3L^2_p}{(1+\frac{2\gamma}{1+4\gamma})}\bigg)+\frac{\gamma
b}{8H^3}\nonumber\\
&\times&a^{-\frac{3\gamma(4k\lambda-1)}{(3k\lambda
\gamma-1)}}\bigg(\frac{(1-4k\lambda)}{(3k\lambda\gamma-1)}+\frac{k\gamma
ba^{-\frac{3\gamma(4k\lambda-1)}{(3k\lambda
\gamma-1)}}}{2H^2}\bigg)\bigg).
\end{eqnarray}
\begin{figure} \centering
\epsfig{file=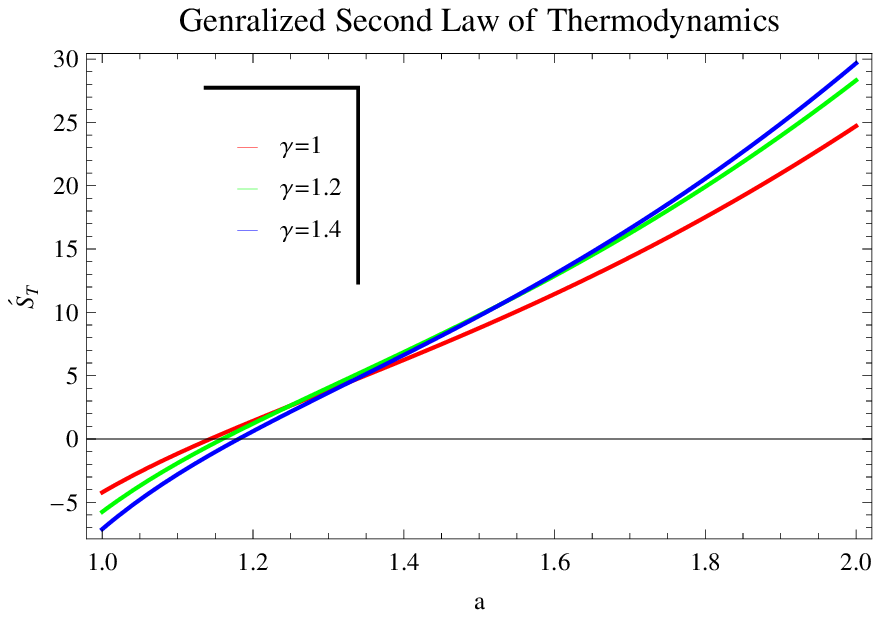,width=.50\linewidth}\caption{Plot of
$\acute{S}_T$ versus $a$ for logarithmic corrected entropy using
$k=1,~\lambda=-1,~\alpha=-1,~L_p=1,~\beta=1$ and $b=1$.}
\end{figure}
In case of logarithmic corrected entropy, we analyze the behavior of
GSLT by plotting the graph of $\acute{S}_T$ versus scale factor as
shown in Figure \textbf{4}. The trajectories of GSLT  meets the
condition $\acute{S}_T\geq0$ for all the three vales of $\gamma$ for
specific ranges of $a$. For $1.26<a,~1.27<a$ and $1.28<a$
corresponding to $\gamma=1,~1.2$ and $1.4$ respectively indicates
the positive behavior expressing the validity of GSLT.

In order to discuss the thermodynamical equilibrium, we again
differentiate the above equation. It is given by
\begin{eqnarray}\label{E23}
S^{''}_T&=&\frac{\sqrt{3}}{a\sqrt{\frac{bk(3k\lambda\gamma-1)a^{-\frac{3\gamma(4k\lambda-1)}{(3k\lambda
\gamma-1)}}}{(4k\lambda-1)}}}\bigg(\frac{27b^2\gamma^2k\sqrt{3}a^{-\frac{3\gamma(4k\lambda-1)}{(3k\lambda
\gamma-1)}}\big(\frac{3\gamma(4k\lambda-1)}{2(3k\lambda\gamma-1)}
-\frac{4k\lambda-1}{3k\lambda\gamma-1}\big)}{16\big(\frac{bk(3k\lambda\gamma-1)a^{-\frac{3\gamma(4k\lambda-1)}{(3k\lambda
\gamma-1)}}}{(4k\lambda-1)}\big)^\frac{5}{2}}\nonumber\\
&-&\frac{9b\gamma^2\sqrt{3}(4k\lambda-1)a^{-1-\frac{3\gamma(4k\lambda-1)}{(3k\lambda
\gamma-1)}}\big(\frac{3\gamma(4k\lambda-1)}{2(3k\lambda\gamma-1)}
-\frac{4k\lambda-1}{3k\lambda\gamma-1}\big)}{8(3k\lambda\gamma-1)
\big(\frac{bk(3k\lambda\gamma-1)a^{-\frac{3\gamma(4k\lambda-1)}{(3k\lambda
\gamma-1)}}}{(4k\lambda-1)}\big)^\frac{3}{2}}+\frac{3\gamma(4k\lambda-1)}{2(3k\lambda\gamma-1)}\nonumber\\
&\times&\bigg(\frac{3bk\gamma\sqrt{3}a^{-1-\frac{3\gamma(4k\lambda-1)}{(3k\lambda
\gamma-1)}}(1+\frac{2\gamma}{1+4\gamma})}{8L^2_p\big(\frac{bk(3k\lambda\gamma-1)a^{-\frac{3\gamma(4k\lambda-1)}{(3k\lambda
\gamma-1)}}}{(4k\lambda-1)}\big)^\frac{3}{2}}-\frac{bk\alpha\gamma\sqrt{3}a^{-1-\frac{3\gamma(4k\lambda-1)}{(3k\lambda
\gamma-1)}}}{\sqrt{\frac{bk(3k\lambda\gamma-1)a^{-\frac{3\gamma(4k\lambda-1)}{(3k\lambda
\gamma-1)}}}{(4k\lambda-1)}}}+8bkL^2_p\beta\nonumber\\
&\times&\gamma a^{-1-\frac{3\gamma(4k\lambda-1)}{(3k\lambda
\gamma-1)}}\sqrt{\frac{bk(3k\lambda\gamma-1)a^{-\frac{3\gamma(4k\lambda-1)}{(3k\lambda
\gamma-1)}}}{(4k\lambda-1)}}\bigg)\bigg)+3bk\gamma\sqrt{3}a^{-2-\frac{3\gamma(4k\lambda-1)}{(3k\lambda
\gamma-1)}}\nonumber\\
&\times&\frac{1}{2\big(\frac{bk(3k\lambda\gamma-1)a^{-\frac{3\gamma(4k\lambda-1)}{(3k\lambda
\gamma-1)}}}{(4k\lambda-1)}\big)^\frac{3}{2}}\bigg(\frac{3b\gamma\sqrt{3}a^{-\frac{3\gamma(4k\lambda-1)}{(3k\lambda
\gamma-1)}}\big(\frac{3\gamma(4k\lambda-1)}{2(3k\lambda\gamma-1)}
-\frac{4k\lambda-1}{3k\lambda\gamma-1}\big)}{8\big(\frac{bk(3k\lambda\gamma-1)a^{-\frac{3\gamma(4k\lambda-1)}{(3k\lambda
\gamma-1)}}}{(4k\lambda-1)}\big)^\frac{3}{2}}\nonumber\\
&+&\frac{3\gamma(4k\lambda-1)}{2(3k\lambda\gamma-1)}\bigg(\frac{\sqrt{3}(1+\frac{2\gamma}{1+4\gamma})}
{4L^2_p\sqrt{\frac{bk(3k\lambda\gamma-1)a^{-\frac{3\gamma(4k\lambda-1)}{(3k\lambda
\gamma-1)}}}{(4k\lambda-1)}}}+\frac{2\alpha\sqrt{\frac{bk(3k\lambda\gamma-1)a^{-\frac{3\gamma(4k\lambda-1)}{(3k\lambda
\gamma-1)}}}{(4k\lambda-1)}}}{\sqrt{3}}\nonumber\\
&-&\frac{16L^2_p\beta\big(\frac{bk(3k\lambda\gamma-1)a^{-\frac{3\gamma(4k\lambda-1)}{(3k\lambda
\gamma-1)}}}{(4k\lambda-1)}\big)^\frac{3}{2}}{3\sqrt{3}}\bigg)\bigg)
-\frac{\sqrt{3}}{a^2\sqrt{\frac{bk(3k\lambda\gamma-1)a^{-\frac{3\gamma(4k\lambda-1)}{(3k\lambda
\gamma-1)}}}{(4k\lambda-1)}}}\nonumber\\
&\times&\bigg(\frac{3b\gamma
a^{-\frac{3\gamma(4k\lambda-1)}{(3k\lambda
\gamma-1)}}\big(\frac{3\gamma(4k\lambda-1)}{2(3k\lambda\gamma-1)}
-\frac{4k\lambda-1}{3k\lambda\gamma-1}\big)}{8\big(\frac{bk(3k\lambda\gamma-1)a^{-\frac{3\gamma(4k\lambda-1)}{(3k\lambda
\gamma-1)}}}{(4k\lambda-1)}\big)^\frac{3}{2}}+\frac{3\gamma(4k\lambda-1)}{2(3k\lambda\gamma-1)}
\bigg(-\frac{16L^2_p\beta}{3\sqrt{3}}\nonumber\\
&\times&\big(\frac{bk(3k\lambda\gamma-1)a^{-\frac{3\gamma(4k\lambda-1)}{(3k\lambda
\gamma-1)}}}{(4k\lambda-1)}\big)^\frac{3}{2}+\frac{\sqrt{3}(1+\frac{2\gamma}{1+4\gamma})}{4L^2_p\sqrt{\frac{bk(3k\lambda\gamma-1)
a^{-\frac{3\gamma(4k\lambda-1)}{(3k\lambda
\gamma-1)}}}{(4k\lambda-1)}}}+\frac{2\alpha}{\sqrt{3}}\nonumber\\
&\times&\sqrt{\frac{bk(3k\lambda\gamma-1)a^{-\frac{3\gamma(4k\lambda-1)}{(3k\lambda
\gamma-1)}}}{(4k\lambda-1)}}\bigg)\bigg).
\end{eqnarray}
\begin{figure} \centering
\epsfig{file=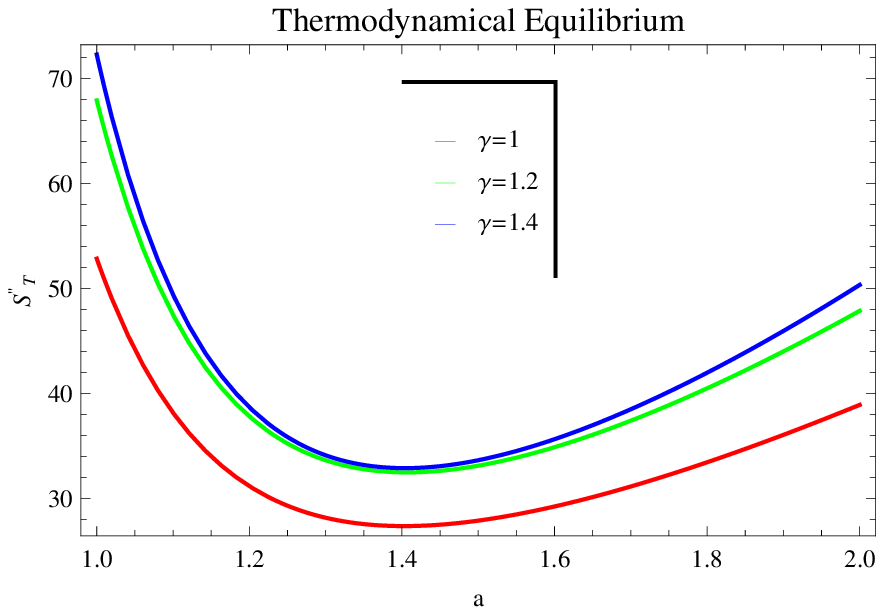,width=.70\linewidth}\caption{Plot of
$S^{''}_T$ versus $a$ for logarithmic corrected entropy using
$k=1,~\lambda=-1,~\alpha=-1,~L_p=1,~\beta=1$ and $b=1$.}
\end{figure}
\begin{figure} \centering
\epsfig{file=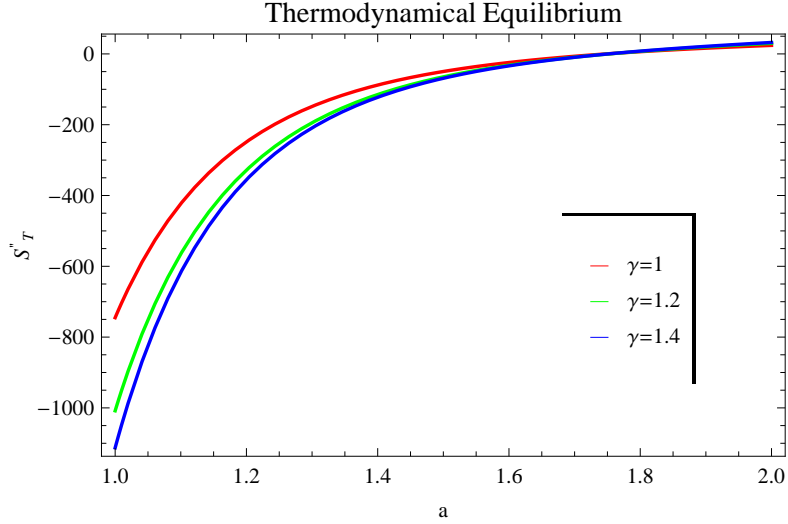,width=.70\linewidth}\caption{Plot of
$S^{''}_T$ versus $a$ for logarithmic corrected entropy using
$k=1,~\lambda=-1,~\alpha=-1,~L_p=1,~b=1$ and $\beta=-1$ in outer
graph while $\beta=-20$ in inner graph.}
\end{figure}
The graphical behavior of $S^{''}_T$ versus $a$ is shown in Figure
\textbf{5} for same constant values as mentioned above. We observe
that all the trajectories express the positive behavior which
represent non-equilibrium state of the solution. However, if we
replace $\beta=-1$, we obtain the equilibrium states for $a\leq
1.2,~ a\leq 1.225,~ a\leq1.24$ for logarithmic corrected entropy
related to $\gamma=1,~1.2,~1.4$ respectively as shown in Figure
\textbf{6} (outer graph). The inner plots in this Figure show the
trajectories for replacing the value of $\beta=-20$ which indicate
the negative behavior for more values of $a$. This leads to the
result that we obtain the thermodynamical equilibrium as we decrease
the value of $\beta$. However, first law of thermodynamics does not
hold while GSLT satisfies for these negative values.

\section{Power-Law Corrected Entropy}

The quantum corrections provided to the entropy-area relationship
lead to the curvature corrections in the Einstein-Hilbert action and
vice versa \cite{14}-\cite{142}. As it has been shown in Ref.
\cite{hooman}, the linear entropy-area relation $(S\sim A)$ in the
Rastall theory is the same as that of the Einstein theory
\cite{ser}. In addition, the entanglement of quantum fields between
inside and outside of the horizon produces an entropy as $A^m$,
where $m$ depends on the amount of mixing \cite{15}. Thus, by adding
this entropy to the horizon entropy \cite{15}, one may get the
power-law corrected entropy as \cite{15}
\begin{eqnarray}\label{E24}
S_{\textmd{A}}=\frac{\tilde{A}}{4L^2_p}\big(1-F_\alpha
\tilde{A}^{1-\frac{\alpha}{2}}\big),
\end{eqnarray}
where
$F_\alpha=\frac{\alpha(4\pi)^{\frac{\alpha}{2}-1}}{(4-\alpha)r^{4-\alpha}_c},~\alpha
$ is a dimensionless constant and $r_c$ represents the crossover
scale. The differential of  Eq.(\ref{E24}) is given by
\begin{eqnarray}\label{E25}
dS_A=\frac{k\gamma ba^{-\frac{3\gamma(4k\lambda-1)}{(3k\lambda
\gamma-1)}}}{2H^2}\bigg(\frac{(1+\frac{2\gamma}{1+4\gamma})}{4HL^2_p}-\frac{(1+\frac{2\gamma}
{1+4\gamma})^{2-\frac{\alpha}{2}}F_\alpha}{4L^2_p}
\big(2-\frac{\alpha}{2}\big)\big(\frac{1}{H}\big)^{3-\alpha}\bigg)dt,
\end{eqnarray}
which leads to
\begin{eqnarray}\label{E26}
T_AdS_A=\frac{2bk\gamma a^{-\frac{3\gamma(4k\lambda-1)}{(3k\lambda
\gamma-1)}}}{H}\bigg(\frac{(1+\frac{2\gamma}{1+4\gamma})}{4HL^2_p}-\frac{F_\alpha}{4L^2_p}
\big(2-\frac{\alpha}{2}\big)\big(\frac{1}{H}\big)^{3-\alpha}\bigg)dt.
\end{eqnarray}
Combining Eqs.(\ref{E14}) and (\ref{E24}), we get
\begin{eqnarray}\label{E26a}
\Omega=\frac{2bk\gamma a^{-\frac{3\gamma(4k\lambda-1)}{(3k\lambda
\gamma-1)}}}{H}\bigg(\frac{(1+\frac{2\gamma}{1+4\gamma})}{4HL^2_p}-\frac{(1+\frac{2\gamma}{1+4\gamma})^{2-\frac{\alpha}{2}}F_\alpha}{4L^2_p}
\big(2-\frac{\alpha}{2}\big)\big(\frac{1}{H}\big)^{3-\alpha}\bigg)-\frac{\gamma
ba^{-\frac{3\gamma(4k\lambda-1)}{(3k\lambda \gamma-1)}}}{2H^2}.
\end{eqnarray}
\begin{figure} \centering
\epsfig{file=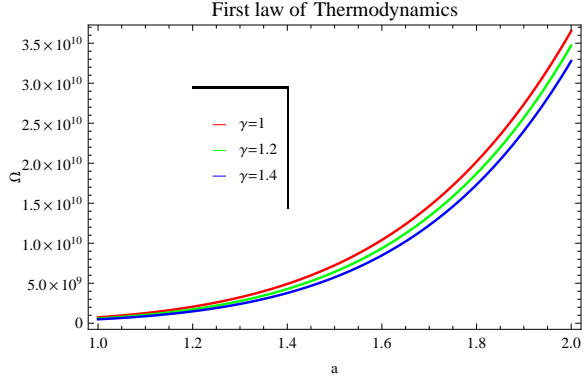,width=.50\linewidth}\caption{Plot of
$\Omega$ versus $a$ for power law corrected entropy using
$k=1,~\lambda=-1,~\alpha=1,~L_p=1,~r_c=\frac{1}{67}$ and $b=1$.}
\end{figure}
Figure \textbf{7} represents that the trajectories of $\Omega$
against $a$ with respect to three values of $\gamma$ approach to
zero which indicates the validity of first law of thermodynamics.

To discuss the GSLT for power law corrected entropy at event
horizon, we obtain the total entropy by using Eqs.(\ref{E18}) and
(\ref{E25}) as
\begin{eqnarray}\label{E27}
\dot{S_T}&=&\frac{\gamma ba^{-\frac{3\gamma(4k\lambda-1)}{(3k\lambda
\gamma-1)}}}{8H^3}\bigg(\frac{(1-4k\lambda)}{(3k\lambda\gamma-1)}+\frac{k\gamma
ba^{-\frac{3\gamma(4k\lambda-1)}{(3k\lambda
\gamma-1)}}}{2H^2}\bigg)+\frac{k\gamma
ba^{-\frac{3\gamma(4k\lambda-1)}{(3k\lambda
\gamma-1)}}}{2H^2}\nonumber\\
&\times&\bigg(\frac{(1+\frac{2\gamma}{1+4\gamma})}{4HL^2_p}
-\frac{(1+\frac{2\gamma}{1+4\gamma})^{2-\frac{\alpha}{2}}F_\alpha}{4L^2_p}
\big(2-\frac{\alpha}{2}\big)\big(\frac{1}{H}\big)^{3-\alpha}\bigg).
\end{eqnarray}
The above equation reduces to
\begin{eqnarray}\label{E27a}
\acute{S}_T&=&\frac{1}{a H}\bigg(\frac{\gamma
ba^{-\frac{3\gamma(4k\lambda-1)}{(3k\lambda
\gamma-1)}}}{8H^3}\bigg(\frac{(1-4k\lambda)}{(3k\lambda\gamma-1)}+\frac{k\gamma
ba^{-\frac{3\gamma(4k\lambda-1)}{(3k\lambda
\gamma-1)}}}{2H^2}\bigg)\nonumber\\
&+&\frac{k\gamma ba^{-\frac{3\gamma(4k\lambda-1)}{(3k\lambda
\gamma-1)}}}{2H^2}\big(\frac{(1+\frac{2\gamma}{1+4\gamma})}
{4HL^2_p}-\frac{(1+\frac{2\gamma}{1+4\gamma})^{2-\frac{\alpha}{2}}F_\alpha}{4L^2_p}
\big(2-\frac{\alpha}{2}\big)\big(\frac{1}{H}\big)^{3-\alpha}\big)\bigg).
\end{eqnarray}
\begin{figure} \centering
\epsfig{file=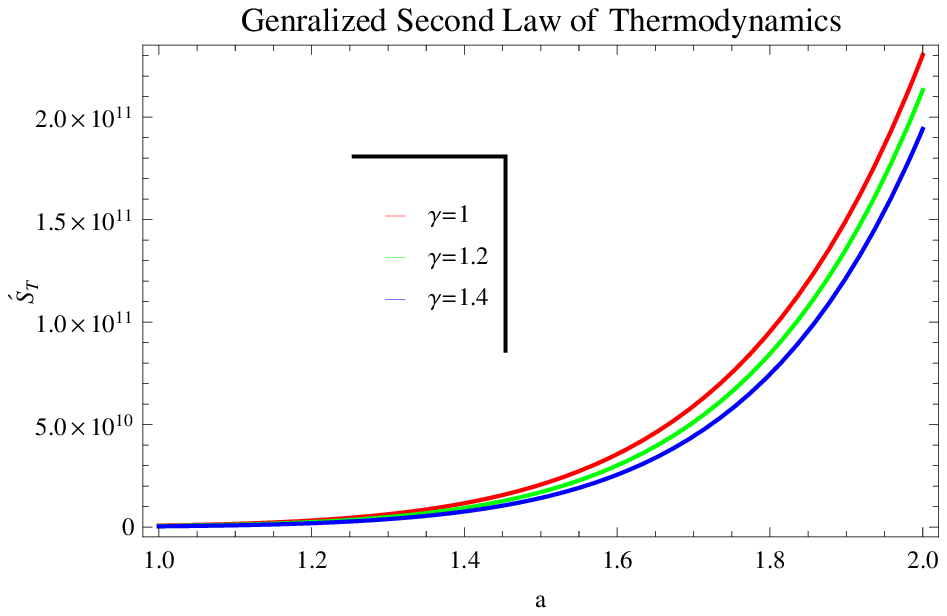,width=.50\linewidth}\caption{Plot of
$\acute{S}_T$ versus $a$ for power law corrected entropy using
$k=1,~\lambda=-1,~\alpha=-1,~L_p=1,~r_c=\frac{1}{67}$ and $b=1$.}
\end{figure}
\begin{figure} \centering
\epsfig{file=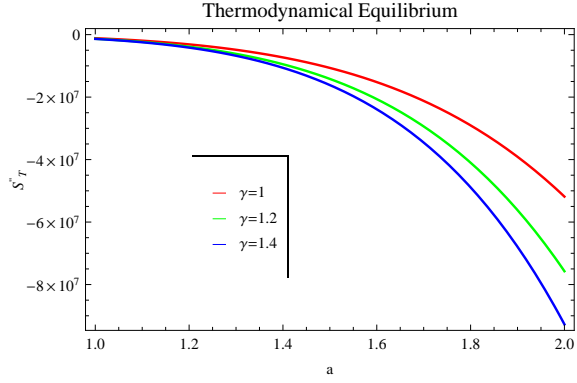,width=.50\linewidth}\caption{Plot of
$S^{''}_T$ versus $a$ for power law corrected entropy using $
k=1,~\lambda=-1,~\alpha=1,~L_p=1,~r_c=\frac{1}{67}$ and $b=1$.}
\end{figure}
The graphical behavior of $\acute{S}_T$ against $a$ is shown in
Figure \textbf{8} for three values of $\gamma$. The trajectories
follow the condition $\acute{S}_T\geq0$ by expressing positive
behavior, which leads to the validity of GSLT for all values
$\gamma$. From Eq.(\ref{E27}), we find the second order differential
equation as follows
\begin{eqnarray}\label{E28}
S^{''}_T&=&\frac{27\gamma^2(4k\lambda-1)^3a^{-1+\frac{3\gamma(4k\lambda-1)}{(3k\lambda
\gamma-1)}}\big(\frac{3\gamma(4k\lambda-1)}{2(3k\lambda\gamma-1)}
-\frac{4k\lambda-1}{3k\lambda\gamma-1}\big)}{16bk^2(3k\lambda\gamma-1)^3}
\bigg(3bk\gamma a^{-1-\frac{3\gamma(4k\lambda-1)}{(3k\lambda
\gamma-1)}}\nonumber\\
&\times&\frac{\sqrt{3}(1+\frac{2\gamma}{1+4\gamma})}{8L^2_p\big(\frac{bk(3k\lambda\gamma-1)a^{-\frac{3\gamma(4k\lambda-1)}{(3k\lambda
\gamma-1)}}}{(4k\lambda-1)}\big)^\frac{3}{2}}+3^{1+\frac{3-\alpha}{2}}(\alpha-3)(2-\frac{\alpha}{2})
a^{-1-\frac{3\gamma(4k\lambda-1)}{(3k\lambda\gamma-1)}}\nonumber\\
&\times&\frac{\alpha bk\gamma
F_\alpha(1+\frac{2\gamma}{1+4\gamma})^{2-\frac{\alpha}{2}}\big(\frac{bk(3k\lambda\gamma-1)a^{-\frac{3\gamma(4k\lambda-1)}{(3k\lambda
\gamma-1)}}}{(4k\lambda-1)}\big)^{-1+\frac{\alpha-3}{2}}}{8L^2_p}\bigg)
+27\gamma^2(4k\lambda-1)^3\nonumber\\
&\times&\frac{a^{-2+\frac{3\gamma(4k\lambda-1)}{(3k\lambda
\gamma-1)}}\big(\frac{3\gamma(4k\lambda-1)}{2(3k\lambda\gamma-1)}
-\frac{4k\lambda-1}{3k\lambda\gamma-1}\big)
\big(-1+\frac{3\gamma(4k\lambda-1)}{3k\lambda\gamma-1}\big)}{16bk^2(-1+3k\lambda\gamma)^3}
\bigg(\frac{{1}}{\sqrt{\frac{bk(3k\lambda\gamma-1)a^{-\frac{3\gamma(4k\lambda-1)}{(3k\lambda
\gamma-1)}}}{(4k\lambda-1)}}}\nonumber\\
&\times&\frac{\sqrt{3}}{4L^2_p}-\frac{F_\alpha3^{\frac{3-\alpha}{2}}(2-\frac{\alpha}{2})
\big(\frac{bk(3k\lambda\gamma-1)a^{-\frac{3\gamma(4k\lambda-1)}{(3k\lambda
\gamma-1)}}}{(4k\lambda-1)}\big)^\frac{\alpha-3}{2}}{4L^2_p}\bigg).
\end{eqnarray}
Figure \textbf{9} is showing the graph of $S^{''}_T$ versus scale
factor for $\alpha=1$ respectively. The graph indicates the
thermodynamical equilibrium as the plots are negative for all the
values of $\gamma$.

\section{The Renyi entropy}

A novel sort of the Renyi entropy has been inspected in various
cosmological and gravitational setups \cite{17,18,non20}. In which
not exclusively is the logarithmic corrected entropy of the
original, the Renyi entropy is utilized based on the fact that the
Bekenstein Hawking entropy $S_{BH}$ is a Tsallis entropy $S_A$
\cite{50}. One can obtain the Renyi entropy $S_R$ \cite{18}
\begin{eqnarray}\label{E34}
S_R=\frac{\ln(1+\eta S_A)}{\eta}.
\end{eqnarray}
The differential of this surface entropy is given by
\begin{eqnarray}\label{E35}
dS_R=\frac{k\gamma ba^{-\frac{3\gamma(4k\lambda-1)}{(3k\lambda
\gamma-1)}}}{2H}\big(\frac{(1+\frac{2\gamma}{4\gamma-1})}{\eta(1+\frac{2\gamma}{4\gamma-1})+8H^2}\big)dt,
\end{eqnarray}
which leads to
\begin{eqnarray}\label{E36}
T_AdS_R=k\gamma ba^{-\frac{3\gamma(4k\lambda-1)}{(3k\lambda
\gamma-1)}}\big(\frac{2(1+\frac{2\gamma}{4\gamma-1})}{\eta(1+\frac{2\gamma}{4\gamma-1})+8H^2}\big)dt.
\end{eqnarray}
Both of these equations take the form
\begin{eqnarray}\label{E36a}
\Omega=k\gamma ba^{-\frac{3\gamma(4k\lambda-1)}{(3k\lambda
\gamma-1)}}\big(\frac{2(1+\frac{2\gamma}{4\gamma-1})}{\eta(1+\frac{2\gamma}{4\gamma-1})+8H^2}\big)-\frac{\gamma
ba^{-\frac{3\gamma(4k\lambda-1)}{(3k\lambda \gamma-1)}}}{2H^2}.
\end{eqnarray}
\begin{figure} \centering
\epsfig{file=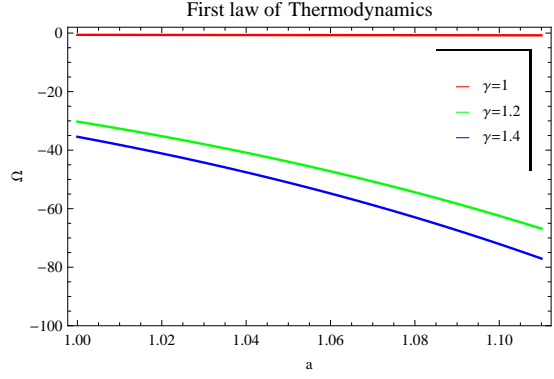,width=.50\linewidth}\caption{Plot of
$\Omega$ versus $a$ for Renyi entropy using
$k=1,~\lambda=-1,~\eta=1$ and $b=1$.}
\end{figure}
The numerical display of above differential equation for  $\Omega$
against $a$ for different values of $\gamma$ is shown in Figure
\textbf{10}. The first law of thermodynamics does not hold  for
$\gamma-1.2,~1.4$ as all the corresponding trajectories fail to meet
the condition $\Omega\rightarrow 0$. The trajectory for $\gamma=1$
represents the validity of first law of thermodynamics. Further, we
analyze the validity of GSLT and thermodynamical equilibrium in the
presence of Renyi entropy. Using Eqs.(\ref{E18}) and (\ref{E35}), we
get
\begin{eqnarray}\label{E37}
\dot{S_T}&=&\bigg(\frac{\gamma
ba^{-\frac{3\gamma(4k\lambda-1)}{(3k\lambda
\gamma-1)}}}{8H^3}\bigg(\frac{(1-4k\lambda)}{(3k\lambda\gamma-1)}+\frac{k\gamma
ba^{-\frac{3\gamma(4k\lambda-1)}{(3k\lambda
\gamma-1)}}}{2H^2}\bigg)\nonumber\\
&+&\frac{(1+\frac{2\gamma}{4\gamma-1})k\gamma
ba^{-\frac{3\gamma(4k\lambda-1)}{(3k\lambda
\gamma-1)}}}{2H}\big(\frac{2H}{\eta(1+\frac{2\gamma}{4\gamma-1})+8H^2}\big)\bigg),
\\\label{E37a} \acute{S}_T&=&\frac{1}{a H}\bigg(\frac{\gamma
ba^{-\frac{3\gamma(4k\lambda-1)}{(3k\lambda
\gamma-1)}}}{8H^3}\bigg(\frac{(1-4k\lambda)}{(3k\lambda\gamma-1)}+\frac{k\gamma
ba^{-\frac{3\gamma(4k\lambda-1)}{(3k\lambda
\gamma-1)}}}{2H^2}\bigg)\nonumber\\
&+&\frac{(1+\frac{2\gamma}{4\gamma-1})k\gamma
ba^{-\frac{3\gamma(4k\lambda-1)}{(3k\lambda
\gamma-1)}}}{2H}\big(\frac{2H}{\eta(1+\frac{2\gamma}{4\gamma-1})+8H^2}\big)\bigg)\bigg).
\end{eqnarray}
\begin{figure} \centering
\epsfig{file=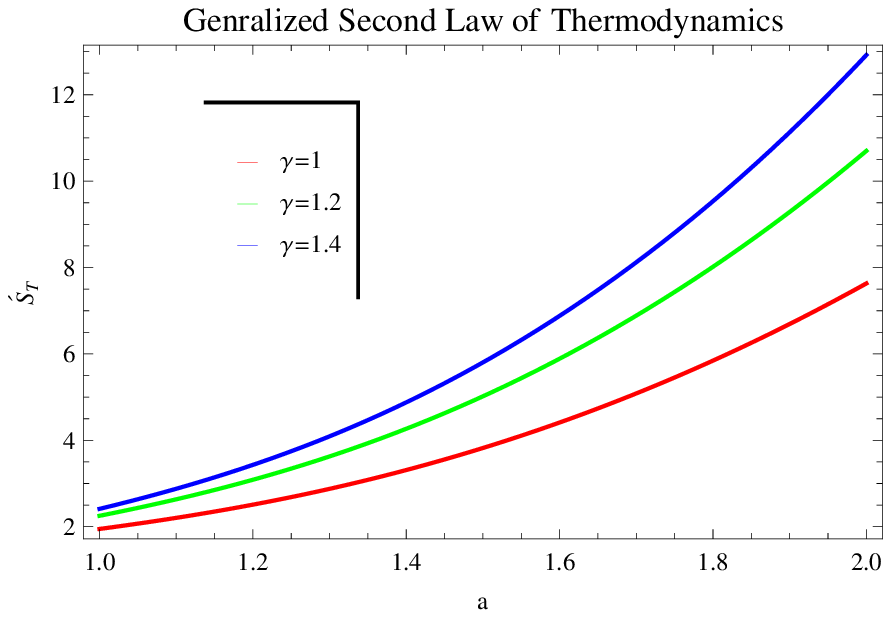,width=.50\linewidth}\caption{Plot of
$\acute{S}_T$ versus $a$ for Renyi entropy using
$k=1,~\lambda=-1,~\eta=1$ and $b=1$.}
\end{figure}
Figure \textbf{11} indicates the plot of $\acute{S}_T$ against scale
factor $a$ for three values of $\gamma$. The trajectories in the
plot are remain positive and obey the condition $\acute{S}_T\geq0$
for all values of $\gamma$ which give the validity of GSLT. The
second order differential equation takes the form
\begin{eqnarray}\label{E39}
S^{''}_T&=&\frac{3\sqrt{3}bk\gamma
a^{-2-\frac{3\gamma(4k\lambda-1)}{(3k\lambda
\gamma-1)}}}{2\big(\frac{bk(3k\lambda\gamma-1)a^{-\frac{3\gamma(4k\lambda-1)}{(3k\lambda
\gamma-1)}}}{(4k\lambda-1)}\big)^\frac{3}{2}}\bigg(\frac{3\sqrt{3}b\gamma
a^{-\frac{3\gamma(4k\lambda-1)}{(3k\lambda
\gamma-1)}}\big(\frac{3\gamma(4k\lambda-1)}{2(3k\lambda\gamma-1)}
-\frac{4k\lambda-1}{3k\lambda\gamma-1}\big)}{8\big(\frac{bk(3k\lambda\gamma-1)a^{-\frac{3\gamma(4k\lambda-1)}{(3k\lambda
\gamma-1)}}}{(4k\lambda-1)}\big)^\frac{3}{2}}\nonumber\\
&+&\frac{bk\gamma(1+\frac{2\gamma}{4\gamma-1})a^{-\frac{3\gamma(4k\lambda-1)}{(3k\lambda
\gamma-1)}}}{\eta(1+\frac{2\gamma}{4\gamma-1})+\frac{8bk(3k\lambda\gamma-1)a^{-\frac{3\gamma(4k\lambda-1)}{(3k\lambda
\gamma-1)}}}{3(4k\lambda-1)}}\bigg)
-\frac{\sqrt{3}}{a^2\sqrt{\frac{bk(3k\lambda\gamma-1)
a^{-\frac{3\gamma(4k\lambda-1)}{(3k\lambda
\gamma-1)}}}{(4k\lambda-1)}}}\nonumber\\
&\times&\bigg(\frac{3\sqrt{3}b\gamma
a^{-\frac{3\gamma(4k\lambda-1)}{(3k\lambda
\gamma-1)}}\big(\frac{3\gamma(4k\lambda-1)}{2(3k\lambda\gamma-1)}
-\frac{4k\lambda-1}{3k\lambda\gamma-1}\big)}{8\big(\frac{bk(3k\lambda\gamma-1)a^{-\frac{3\gamma(4k\lambda-1)}{(3k\lambda
\gamma-1)}}}{(4k\lambda-1)}\big)^\frac{3}{2}}+bk\gamma(1+\frac{2\gamma}{4\gamma-1})
a^{-\frac{3\gamma(4k\lambda-1)}{(3k\lambda
\gamma-1)}}\nonumber\\
&\times&\frac{1}{\eta(1+\frac{2\gamma}{4\gamma-1})+\frac{8bk(3k\lambda\gamma-1)
a^{-\frac{3\gamma(4k\lambda-1)}{(3k\lambda
\gamma-1)}}}{3(4k\lambda-1)}}\bigg)+\frac{\sqrt{3}}{a\sqrt{\frac{bk(3k\lambda\gamma-1)a^{-\frac{3\gamma(4k\lambda-1)}{(3k\lambda
\gamma-1)}}}{(4k\lambda-1)}}}\nonumber\\
&\times&\bigg(\frac{27b^2\gamma^2k\sqrt{3}a^{-1-\frac{6\gamma(4k\lambda-1)}{(3k\lambda
\gamma-1)}}\big(\frac{3\gamma(4k\lambda-1)}{2(3k\lambda\gamma-1)}
-\frac{4k\lambda-1}{3k\lambda\gamma-1}\big)}{16\big(\frac{bk(3k\lambda\gamma-1)a^{-\frac{3\gamma(4k\lambda-1)}{(3k\lambda
\gamma-1)}}}{(4k\lambda-1)}\big)^\frac{5}{2}}-9b\gamma^2\sqrt{3}a^{-\frac{3\gamma(4k\lambda-1)}{(3k\lambda
\gamma-1)}}\nonumber\\
&\times&\frac{(4k\lambda-1)\big(\frac{3\gamma(4k\lambda-1)}{2(3k\lambda\gamma-1)}
-\frac{4k\lambda-1}{3k\lambda\gamma-1}\big)}{8(3k\lambda\gamma-1)
\big(\frac{bk(3k\lambda\gamma-1)a^{-\frac{3\gamma(4k\lambda-1)}{(3k\lambda
\gamma-1)}}}{(4k\lambda-1)}\big)^\frac{3}{2}}+\frac{8b^2k^2\gamma^2(1+\frac{2\gamma}{4\gamma-1})
a^{-1-\frac{6\gamma(4k\lambda-1)}{(3k\lambda
\gamma-1)}}}{\big(\eta(1+\frac{2\gamma}{4\gamma-1})+\frac{8bk(3k\lambda
\gamma-1)a^{-\frac{3\gamma(4k\lambda-1)}{(3k\lambda \gamma-1)}}
}{3(4k\lambda-1)}\big)^2}\nonumber\\
&-&\frac{3bk\gamma^2(4k\lambda-1)(1+\frac{2\gamma}{4\gamma-1})a^{-1-\frac{3\gamma(4k\lambda-1)}{(3k\lambda
\gamma-1)}}}{(3k\lambda\gamma-1)\big(\eta(1+\frac{2\gamma}{4\gamma-1})+\frac{8bk(3k\lambda\gamma-1)
a^{-\frac{3\gamma(4k\lambda-1)}{(3k\lambda
\gamma-1)}}}{3(4k\lambda-1)}\big)}\bigg)
\end{eqnarray}
\begin{figure} \centering
\epsfig{file=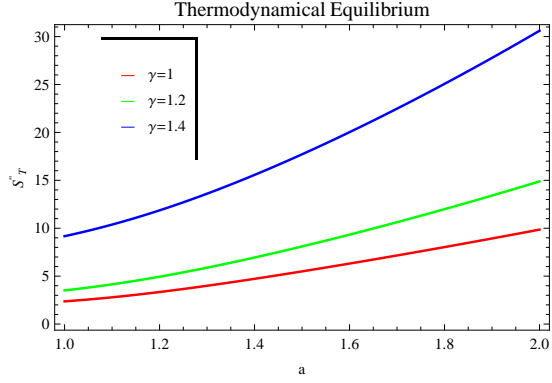,width=.50\linewidth}\caption{Plot of
$S^{''}_T$ versus $a$ for Renyi entropy using
$k=1,~\lambda=-1,~\eta=1$ and $b=1$.}
\end{figure}
The plot of $S^{''}_T$ versus $a$ for second order differential
equation with apparent horizon as shown in Figure\textbf{12}. It is
observed that $S^{''}_T\geq0$  with all values of
$\frac{2}{3}\leq\gamma\leq 2$ which leads to instability of
thermodynamical equilibrium with $S^{''}_T<0$.

\section{Conclusions}

In the present paper, we have investigated the validity of first
law of thermodynamics, GSLT and thermodynamical equilibrium for
the flat FRW universe in Rastall theory of gravity. For this
purpose, we have taken the EoS for perfect fluid by considering
the different entropies including the modified Bekenstein entropy,
the logarithmic corrected entropy, power law corrected entropy and
the Renyi entropy. We have summarized our results as follows:
\begin{itemize}
\item \textbf{For the modified Bekenstein entropy}
The plot of $\acute{S}_T$ versus scale factor parameter as shown in
Figure \textbf{1} prove that GSLT is valid for all values of
$\frac{2}{3}\leq\gamma\leq 2$. Further, we have observed the
validity of thermodynamical equilibrium. Figure \textbf{2} indicates
that thermodynamical equilibrium satisfies the condition
$S^{''}_T<0$.
 \item \textbf{For Logarithmic corrected entropy}
In the presence of logarithmic  corrected entropy it can be seen
that first law of thermodynamics is showing the validity for some
specific points. These are for $\gamma=1$ at $a=1.39$, at $a=1.43$
for $\gamma=1.2$ and for $\gamma=1.4$ at $a=1.45$ represent the
validity of first law of thermodynamics (Figure \textbf{3}). The
trajectories of GSLT meets the condition $\acute{S}_T\geq0$ for all
the three vales of $\gamma$ for specific ranges of $a$ which are
$a>1.07,~a>1.08$ and $a>1.09$ corresponding to $\gamma=1,~1.2$ and
$1.4$ respectively. (Figure \textbf{4}). Further, the graphical
behavior of $S^{''}_T$ against $a$ is shown in Figure \textbf{5}
does not hold the thermodynamical equilibrium when $\beta$ is
positive while Figure \textbf{6} provide the validity of
thermodynamical equilibrium for all values of $\gamma$ with negative
decreasing value of $\beta$.
  \item \textbf{Power law Corrected Entropy}
In this entropy, we have analyzed that the first law of
thermodynamics holds (Figure \textbf{7}) as well as the GSLT is
valid for all values $\gamma$ (Figure \textbf{8}). From Figure
(\textbf{9}), we have investigated that the thermodynamical
equilibrium condition $S^{''}_T<0)$ satisfied with all values $a$
The trajectories of thermodynamical equilibrium are negative which
lead to the instability of thermodynamical equilibrium.
\item \textbf{For the Renyi Entropy}
In this entropy, we have observed that the first law of
thermodynamics does not hold  for $\gamma-1.2,~1.4$ as all the
corresponding trajectories fail to meet the condition
$\Omega\rightarrow 0$. The trajectory for $\gamma=1$ represents the
validity of first law of thermodynamics. (Figure \textbf{10}). The
graphical behavior of Figure \textbf{11} shows that all trajectories
remains positive for all values of $\gamma$ which leads to the
validity of GSLT. Moreover, thermodynamical equilibrium condition is
not satisfied with all values of $\gamma$ (Figure \textbf{12}).
\end{itemize}

\section*{Acknowledgment}

This work was supported in part by the JSPS KAKENHI Grant Number
JP 25800136 and Competitive Research Funds for Fukushima University Faculty (17RI017) (K.B.). The work of H. Moradpour has been supported
financially by Research Institute for Astronomy \& Astrophysics of
Maragha (RIAAM) under research project No. $1/5237-7$.


\begin{thebibliography}{43}

\bibitem{s1} E. Poisson, \emph{A Relativist Toolkit} (Cambridge University Press, UK,
2004).

\bibitem{s2} N. K. Glendenning, Special and General Relativity: With Applications
to White Dwarfs, Neutron Stars and Black Holes, (Springer, USA,
2007).

\bibitem{s3} M. Roos, \emph{Introduction to Cosmology} (John Wiley and Sons, UK, 2003).
\bibitem{1} P. Rastall, Phys. Rev. D \textbf{6}, 3357 (1972).

\bibitem{s3b} G. W. Gibbons and S. W. Hawking, Phys. Rev. D \textbf{15}, 2738 (1977).

\bibitem{s3c} L. Parker, Phys. Rev. D\textbf{3}, 346 (1971); D\textbf{3}, 2546 (1971).

\bibitem{s3d} L. H. Ford, Phys. Rev. D \textbf{35}, 2955 (1987).

\bibitem{s4d} N. D. Birrell, and P. C. W. Davies, \emph{Quantum Fieldsin Curved Space}
(Cambridge University Press, Cambridge, 1982).

\bibitem{s5d}  C. E. M. Batista, M. H. Daouda, J. C. Fabris, O. F. Piattella, and
D. C. Rodrigues, Phys. Rev. D \textbf{85}, 084008 (2012).


\bibitem{R-DE-MG}
%
S.~Nojiri and S.~D.~Odintsov,
Phys.\ Rept.\ {\bf 505}, 59 (2011).

\bibitem{R-DE-MG1}
%
S.~Nojiri and S.~D.~Odintsov,
eConf C {\bf 0602061} (2006) 06 [Int.\ J.\ Geom.\ Meth.\ Mod.\
Phys.\ {\bf 4}, 115 (2007)].
\bibitem{R-DE-MG3}
%
%
S.~Capozziello and V.~Faraoni, \textit{Beyond Einstein Gravity}
(Springer, Dordrecht, 2010).
\bibitem{R-DE-MG4}
%
%
S.~Capozziello and M.~De Laurentis,
Phys.\ Rept.\ {\bf 509}, 167 (2011).
\bibitem{R-DE-MG5}
%
%
%
  K.~Bamba, S.~Capozziello, S.~Nojiri and S.~D.~Odintsov,
  Astrophys.\ Space Sci.\  {\bf 342}, 155 (2012).
  \bibitem{R-DE-MG6}
%
%
  A.~Joyce, B.~Jain, J.~Khoury and M.~Trodden,
  Phys.\ Rept.\  {\bf 568}, 1 (2015).
  \bibitem{R-DE-MG7}
%
%
  K.~Koyama,
  Rept.\ Prog.\ Phys.\  {\bf 79}, 046902 (2016).
  \bibitem{R-DE-MG8}
%
%
  K.~Bamba and S.~D.~Odintsov,
  Symmetry {\bf 7}, 1, 220 (2015).
%
\bibitem{Nojiri:2017ncd}
  S.~Nojiri, S.~D.~Odintsov and V.~K.~Oikonomou,
  Phys.\ Rept.\  {\bf 692}, 1 (2017).
\bibitem{s4} T. Jacobson, Phys. Rev. Lett. \textbf{75}, 1260 (1995).
\bibitem{s5} C. Eling, R. Guedens, and T. Jacobson, Phys. Rev. Lett. \textbf{96}, 121301 (2006).
\bibitem{s6} M. Akbar, and R. G. Cai, Phys. Lett. B \textbf{648}, 243 (2007).
\bibitem{s8} T. Padmanabhan, Phys. Rept. \textbf{406}, 49 (2005).
\bibitem{s12} R. G. Cai, and S. P. Kim, JHEP \textbf{0502}, 050 (2005).
\bibitem{s7} R. G. Cai, and L. M. Cao, Phys. Rev. D \textbf{75}, 064008 (2007).
\bibitem{s9} S. W. Hawking, Phys. Rev. Lett. \textbf{26}, 1344 (1971).
\bibitem{s10} T. Padmanabhan, Rep. Prog. Phys. \textbf{73}, 046901 (2010).
\bibitem{s13} T. Padmanabhan, Class. Quantum Grav. \textbf{19}, 5387 (2002).
\bibitem{j6} A. Sheykhi, B. Wang, R. G. Cai, Nucl. Phys. B {\bf779}, 1 (2007).
\bibitem{j7} A. Sheykhi, B. Wang, R. G. Cai, Phys. Rev. D {\bf76}, 023515 (2007).
\bibitem{j13} A. Sheykhi, M. H. Dehghani, R. Dehghani, Gen. Relativ. Gravit. {\bf46}, 1679 (2014).
\bibitem{msgj} H. Moradpour, N. Sadeghnezhad, S. Ghaffari, and A. Jahan, AHEP. Article ID 9687976, (2017).
\bibitem{plb} H. Moradpour, Phys. Lett. B 757, 187 (2016), final version in arXiv:1601.04529.
\bibitem{ms} H. Moradpour, Ines. G. Salako, AHEP. Article ID 3492796, (2016).
\bibitem{plb1c} F. F. Yuan, P. Huang, arXiv:1607.04383.
\bibitem{s14} J. D. Bekenstein, Phys. Rev. D \textbf{7}, 2333 (1973).
\bibitem{s16} R. Banerjee, and S. K. Modak, JHEP \textbf{073}, 0911 (2009).
\bibitem{s161} H. Wei, Commun. Theor. Phys. \textbf{52}, 743 (2009).
\bibitem{s162} S. Banerjee, R. K. Gupta, and A. Sen, JHEP \textbf{147},
1103 (2011).
\bibitem{s17} A. Sheykhi, and M. Jamil, Gen. Relativ. Gravit. \textbf{43}, 2661 (2011).
\bibitem{Bamba:2012mj}
  K.~Bamba, M.~Jamil, D.~Momeni and R.~Myrzakulov,
  Int.\ J.\ Mod.\ Phys.\ D {\bf 21}, 1250065 (2012).

\bibitem{s18} K. Karami, S. Ghaffari, and M. M. Soltanzadeh, Class. Quantum Grav.
\textbf{27}, 205021 (2010).

\bibitem{s181} M. R. Setare, JCAP \textbf{01}, 023 (2007).

\bibitem{s182} A. Sheykhi, Class. Quantum Grav. \textbf{27}, 025007
(2010).

\bibitem{s183} M. Mazumder, and S. Chakraborty, Gen. Relativ. Gravit. \textbf{42},
813 (2010).

\bibitem{s19} M. Jamil, E. N. Saridakis, and M. R. Setare, Phys. Rev. D
\textbf{81}, 023007 (2010).

\bibitem{s191} K. Karami, et al., JHEP\textbf{150}, 1108 (2011).

\bibitem{s192} K. Karami, et al, Eur. Phys. Lett.\textbf{93}, 29002 (2011).

\bibitem{CT}
%
  K.~Bamba and C.~Q.~Geng,
  Phys.\ Lett.\ B {\bf 679}, 282 (2009).
  \bibitem{CT1}
%
  K.~Bamba, C.~Q.~Geng and S.~Tsujikawa,
  Phys.\ Lett.\ B {\bf 688}, 101 (2010).
%
\bibitem{CT2}
  K.~Bamba, C.~Q.~Geng, S.~Nojiri and S.~D.~Odintsov,
  EPL {\bf 89}, 50003 (2010).
%
\bibitem{CT3}
  K.~Bamba and C.~Q.~Geng,
  JCAP {\bf 1006}, 014 (2010).
%
\bibitem{CT4}
  K.~Bamba and C.~Q.~Geng,
  JCAP {\bf 1111}, 008 (2011).
%
\bibitem{CT5}
  K.~Bamba, R.~Myrzakulov, S.~Nojiri and S.~D.~Odintsov,
  Phys.\ Rev.\ D {\bf 85}, 104036 (2012).
%
\bibitem{CT6}
  K.~Bamba, M.~Jamil, D.~Momeni and R.~Myrzakulov,
  Astrophys.\ Space Sci.\  {\bf 344}, 259 (2013).
%

\bibitem{Bamba:2016aoo}
  K.~Bamba,
  Int.\ J.\ Geom.\ Meth.\ Mod.\ Phys.\  {\bf 13}, 1630007 (2016).


\bibitem{lag1} L. L. Smalley, Il Nuovo Cimento B, {\bf 80}, 1, 42 (1984)
\bibitem{lag2} R. V. dos Santos, J. A. C. Nogales, arXiv:1701.08203v1.
\bibitem{lag3} V. Dzhunushaliev and H. Quevedo, Grav. Cosm. 23, 280 (2017).
\bibitem{lag4} H. Moradpour, I. Licata, C. Corda, Ines G. Salako, arXiv:1802.00738v1.
\bibitem{epjcc0} L. L. Smalley, J. Phys. A: Math. Gen. 16, 2179 (1983).
\bibitem{epjcc} F. Darabi, H. Moradpour, I. Licata, Y. Heydarzade, C. Corda, EPJC 78, 25 (2018).
\bibitem{non20} H. Moradpour, A. Bonilla, E. M. C. Abreu, J. A. Neto, Phys. Rev. D 96, 123504 (2017).
\bibitem{fabun} Mahamadou Daouda, J. C. Fabris, A. M. Oliveira, F. Smirnov, H. E. S. Velten, arXiv:1802.01413v1.

\bibitem{3} M. Roos, Introduction to Cosmology (John Wiley and Sons, UK, 2003).

\bibitem{4} M. Capone, V. F. Cardone, and M. L. Ruggiero, Journal of Physics:
Conference Series, D\textbf{222}, 012012 (2010).

\bibitem{5} L. L. Smalley, and N. Cimento B \textbf{80(1)}, 42 (1984).

\bibitem{6} R. Bousso, Phys. Rev. D \textbf{71}, 064024 (2005).

\bibitem{7} K. A. Miessner, Class. Quantum Grav. \textbf{21}, 5245 (2004).

\bibitem{8} A. Ghosh, and P. Mitra, Phys. Rev. D \textbf{71}, 027502 (2004).
\bibitem{wald} R. M. Wald, Phys. Rev. D 48, 3427 (1993).
\bibitem{090} R. K. Kaul, P. Majumdar, Phys. Rev. Lett. 84, 5255 (2000).
\bibitem{0901} S. Carlip, Class. Quant. Grav. 17, 4175 (2000).
\bibitem{0902} K. Nouicer, Phys. Lett. B 646, 63 (2007).
\bibitem{90} S. Das, P. Majumdar, R. K. Bhaduri, Class. Quant. Grav.19, 2355 (2002).
\bibitem{9} A. Chatterjee, and P. Majumder, Phys. Rev. Lett. \textbf{92}, 141301 (2004).

\bibitem{10} R. Banerjee, and S. K. Modak, JHEP \textbf{0905}, 063 (2009).

\bibitem{11} S. K. Modak, Phys. Lett. B \textbf{671}, 167 (2009).

\bibitem{12} M. Jamil, and M. U. Farooq, JCAP \textbf{03}, 001 (2010).

\bibitem{13} H. M. Sadjadi, and M. Jamil, Europhys. Lett. \textbf{92}, 69001 (2010).
\bibitem{epjcn1} S. Mahapatra, Eur. Phys. J. C 78, 23 (2018).

\bibitem{14} R. Banerjee, and S. K. Modak, JHEP \textbf{05}, 063 (2009).

\bibitem{141} S. K. Modak, Phys. Lett. B \textbf{671}, 167 (2009).

\bibitem{142} R. Banerjee, S. Gangopadhyay, and S. K. Modak, Phys. Lett. B
\textbf{686}, 181 (2010).
\bibitem{ser} M. Srednicki, Phys. Rev. Lett. 71, 666 (1993).
\bibitem{15} S. Das, S. Shankaranarayanan, P. Sur, Phys. Rev. D  \textbf{77}, 064013 (2008).

\bibitem{16} B. Pourhassan, S. Upadhyay, and H. Farahani,
http://arxiv.org/abs/1701.08650.

\bibitem{17} T. S. Biro, and V. G. Czinner,  Phys. Lett. B \textbf{726}, 861 (2013).

\bibitem{18} V. G. Czinner, and H. Iguchi, Phys. Lett. B \textbf{752}, 306 (2016).

\bibitem{Visser:2017gpz}
  M.~Visser,
  Phys.\ Lett.\ B {\bf 782}, 83 (2018)
  doi:10.1016/j.physletb.2018.05.028
  [arXiv:1711.11500 [gr-qc]].

\bibitem{Moradpour:2017tbp}
  F.~Darabi, H.~Moradpour, I.~Licata, Y.~Heydarzade and C.~Corda,
  Eur.\ Phys.\ J.\ C {\bf 78}, 25 (2018)
  doi:10.1140/epjc/s10052-017-5502-5
  [arXiv:1712.09307 [gr-qc]].

\bibitem{hooman} H. Moradpour, I. G. Salako, Adv. High Ener. Phys. \textbf{2016}, 3492796 (2016).
\bibitem{mis} R. G. Cai, L. M. Cao, Y. P. Hu, N. Ohta, Phys. Rev. D 80, 104016 (2009);\\ A. Paranjape, S. Sarkar, T.
Padmanabhan, Phys. Rev. D 74, 104015 (2006).
\bibitem{epjcn} D. Das, S. Dutta, S. Chakraborty, Eur. Phys. J. C 78, 810 (2018).
\bibitem{clas} F. F. Yuan, P. Huang, Class. Quantum Gravity 34, 077001 (2017).
\bibitem{CaiKimt} R. G. Cai, L. M. Cao, Y. P. Hu, Class. Quantum. Grav. {\bf 26}, 155018 (2009).
\bibitem{50} A. Majhi, Phys. Lett. B 775, 32 (2017).

\end{thebibliography}
\end{document}